\pdfoutput=1 
\documentclass[prl,twocolumn,showpacs,superscriptaddress]{revtex4}

\usepackage{amsmath,amssymb}
\usepackage{verbatim}
\usepackage{graphicx}
\usepackage{color}
\usepackage[colorlinks=true,linkcolor=red,citecolor=magenta,urlcolor=blue,filecolor=black]{hyperref}

\DeclareFontFamily{OT1}{rsfs}{}
\DeclareFontShape{OT1}{rsfs}{m}{n}{ <-7> rsfs5 <7-10> rsfs7 <10->rsfs10}{} 
\DeclareMathAlphabet{\mycal}{OT1}{rsfs}{m}{n}

\newcommand{\be}[1]{ \begin{equation}\label{#1} }
\newcommand{\ee}{\end{equation}}
\newcommand{\bea}[1]{\begin{eqnarray}\label{#1} }
\newcommand{\eea}{\end{eqnarray}}

\newcommand{\beqn}{\begin{eqnarray}}
\newcommand{\eeqn}{\end{eqnarray}}
\newcommand{\D}{\text{d}}
\newcommand{\pa}{\partial}
\newcommand{\beq}{\begin{equation}}
\newcommand{\eeq}{\end{equation}}
\newcommand{\ve}{\varepsilon}

\begin{document}

\title{New Chiral Gravity}

\author{Luca Ciambelli}
\email{luca.ciambelli@ulb.ac.be}
\affiliation{Physique Th{\'e}orique et Math{\'e}matique, Universit{\'e} libre de Bruxelles and International Solvay Institutes, Campus Plaine C.P. 231, B-1050 Bruxelles, Belgium}

\author{St\'ephane Detournay}
\email{sdetourn@ulb.ac.be}
\affiliation{Physique Th{\'e}orique et Math{\'e}matique, Universit{\'e} libre de Bruxelles and International Solvay Institutes, Campus Plaine C.P. 231, B-1050 Bruxelles, Belgium}

\author{Antoine Somerhausen}
\email{antoine.somerhausen@ulb.be}
\affiliation{Physique Th{\'e}orique et Math{\'e}matique, Universit{\'e} libre de Bruxelles and International Solvay Institutes, Campus Plaine C.P. 231, B-1050 Bruxelles, Belgium}

\date{\today}

\begin{abstract} 

The phase space of three-dimensional gravity with Compere-Song-Strominger (CSS) boundary conditions is endowed with asymptotic symmetries consisting in the semi-direct product of a Virasoro and a $\hat{u}(1)$ Ka\v{c}-Moody algebra, and contains BTZ black holes whose entropy can be accounted for by the degeneracy of states of a Warped CFT. By embedding these boundary conditions in Topologically Massive Gravity, we observe the existence of two special points in the space of couplings parameterized by the AdS$_3$ radius $\ell$ and the Chern-Simons coupling $\mu$. When $\mu = \pm {1\over \ell}$, the asymptotic symmetries reduce to either a chiral Virasoro algebra or a pure $\hat{u}(1)$ Ka\v{c}-Moody current algebra. At those points, black holes have positive energy while that of linearized excitations are non-negative.

\end{abstract}


\maketitle

\section{Introduction}

Lower-dimensional gravity models have appeared over the years as a fertile playground to address fundamental questions in quantum gravity, such as the black hole entropy problem \cite{Bekenstein:1972tm, Bekenstein:1973ur} of the black hole information paradox \cite{Hawking:1974rv, Hawking:1974sw}. In particular, Einstein-Hilbert gravity in (2+1)-dimensions \cite{Gott:1982qg, Giddings:1983es, Deser:1983tn, Barr} with a negative cosmological constant $\Lambda = -\frac{1}{\ell^2}$ has emerged as a very insightful toy model. Even though the  theory has no bulk propagating degrees of freedom, it has asymptotically AdS$_3$ (BTZ) black hole solutions \cite{Banados:1992wn, Bunster:2014mua}, as well as massless gravitons which can be viewed as propagating on the boundary. Furthermore, the phase space of AdS$_3$ gravity admits a non-trivial action of the 2-dimensional conformal group \cite{Brown:1986nw}, which appeared as the first hint of a deep connection between a gravity theory in AdS space and a CFT in one dimension less, later unravelled by the AdS/CFT correspondence \cite{Maldacena:1997re}. Remarkably, the Bekenstein-Hawking entropy of the BTZ black holes could be reproduced by a counting of states in the corresponding dual CFT \cite{Strominger:1997eq}, opening a new perspectives on the black hole entropy problem. More recently, decisive progress has also been achieved regarding the black hole information paradox relying on a new understanding of how to compute the entropy of Hawking radiation \cite{Penington:2019npb, Almheiri:2019psf}. The key point is the inclusion in the gravitational path integral used to compute the entropy of new saddles arising as complexified wormholes \cite{Almheiri:2019qdq}, which could be explicitly identified in two-dimensional Jackiw-Teitelboim gravity \cite{Teitelboim:1983ux, Jackiw:1984je, Almheiri:2014cka} (which can be viewed as the dimensional reduction of 3d gravity \cite{Mertens:2018fds, Cotler:2018zff}) coupled to matter.

The BTZ entropy derivation relied only on conformal symmetry and few additional assumptions such as unitarity \cite{Cardy:1986ie} and sparseness of the spectrum \cite{Hartman:2014oaa}. The precise nature of the 2d CFT dual to pure gravity remained therefore elusive until 2007 when Witten, assuming holomorphic factorization, proposed that the theory should be an extremal CFT, establishing intriguing connexions between quantum gravity, group and number theory \cite{Witten:2007kt}. Later Maloney and Witten computed the 3d gravity partition function as a sum over topologies and found that the result could not be interpreted as a trace over some CFT Hilbert space \cite{Maloney:2007ud}. Moreover, there are arguments (although no proof) that extremal CFTs could not exist for large central charge, i.e. in the semi-classical regime \cite{Gaberdiel:2007ve, Gaberdiel:2008pr, Gaberdiel:2010jf}. Various steps have been taken towards fixing them \cite{Keller:2014xba, Benjamin:2019stq, Alday:2019vdr, Benjamin:2020mfz}, but it is still unclear whether pure 3d gravity could make sense as a quantum theory. Few months after Witten's proposal, Li, Song and Strominger suggested an alternative for a fully consistent and unitary gravity theory with partition function that of an extremal CFT under the name {\itshape chiral gravity} \cite{Li:2008dq}. By supplementing Einstein gravity with a gravitational Chern-Simons term, they argued that the resulting theory -- Topologically Massive Gravity (TMG) \cite{Deser:1981wh, Deser:1982vy} -- becomes chiral at a specific point in coupling space. Heated debates ensued as to determine whether the theory was actually chiral and unitary (see e.g. \cite{Carlip:2008jk, Grumiller:2008qz, Li:2008yz, Park:2008yy, Grumiller:2008pr, Carlip:2008eq, Carlip:2008qh,  Giribet:2008bw, Strominger:2008dp, Maloney:2009ck, Carlip:2009ey, Skenderis:2009nt, Andrade:2009ae, Henneaux:2010fy, Lashkari:2010iy, Cunliff:2010rv, Nazaroglu:2011zi, Castro:2011ke, Dengiz:2013hka, Altas:2018dci}), leading to unexpected correspondences between critically tuned AdS$_3$ gravity and a logarithmic CFT \cite{Grumiller:2009mw, Gaberdiel:2010xv, Grumiller:2010tj, Grumiller:2013at, Mvondo-She:2018htn}. Whether chiral gravity could exist as a unitary truncation of the non-unitary logarithmic CFT that is dual to TMG at the critical point is still an open question, but still stands out as a candidate for the simplest and potentially solvable model including quantum black holes.

The core of this work is based on the recent observations that pure AdS$_3$ ought not to be dual to 2d CFT in the first place. In defining the classical phase space of a physical theory, boundary conditions play as an important role as the action or the equations of motion. It was noticed in a series of works that besides the classic Brown-Henneaux boundary conditions with conformal symmetry, a variety of alternative consistent choices could exist \cite{Compere:2013bya, Troessaert:2013fma,  Avery:2013dja, Troessaert:2015gra, Donnay:2015abr, Perez:2016vqo, Afshar:2016uax, Afshar:2016kjj, Grumiller:2017sjh}. In particular, it was observed by Compere, Song and Strominger (CSS) that pure 3d gravity with appropriate boundary conditions (referred to as CSS) has asymptotic symmetries consisting in the semi-direct product of a Virasoro and a $\hat{u}(1)$ Ka\v{c}-Moody algebra. Those symmetries are those of a new type of 2d field theories, Warped CFTs (WCFTs) \cite{Hofman:2011zj, Detournay:2012pc}. The study of WCFTs was triggered by the search for holographic duals to the near-horizon region of extremal black holes \cite{Bardeen:1999px, Dias:2007nj, Guica:2008mu} and Warped AdS$_3$ (WAdS$_3$) spaces \cite{Banados:2005da, Anninos:2008fx}, and they have attracted a lot of attention in recent years \cite{Compere:2007in, Compere:2008cv, Compere:2009zj, Blagojevic:2009ek, Henneaux:2011hv, Compere:2013bya, Hofman:2014loa, Castro:2015uaa, Castro:2015csg, Song:2016pwx, Song:2016gtd,  Song:2017czq, Castro:2017mfj, Jensen:2017tnb, Azeyanagi:2018har, Aggarwal:2019iay, Apolo:2020bld, Apolo:2020qjm}. The main goal of the present work we will be to show that embedding CSS boundary conditions in TMG allows to identify two special points in coupling space where the warped symmetries either reduce to a chiral Virasoro algebra or to a $\hat{u}(1)$ Ka\v{c}-Moody algebra, thereby possibly providing the simplest example of a gravity theory including black holes.

 \section{CSS boundary conditions}

The general solution of Einstein equations with CSS boundary conditions is given by \cite{Compere:2013bya}
\beqn
\D s^{2} &=& \ell^{2} \frac{\D \rho^{2}}{\rho^{2}}- \rho^{2} \D x^+\left(\D x^--\partial_{+} \bar{P} \D x^+\right)
\nonumber\\
&&
+4 G \ell\left[\bar{L} \D x^{+2}+\Delta\left(\D x^--\partial_{+} \bar{P} \D x^+\right)^{2}\right] \nonumber\\
&&-\frac{16 G^{2}\ell^2}{\rho ^{2}} \Delta \bar{L} \D x^+\left(\D x^--\partial_{+} \bar{P} \D x^+\right),
 \label{SolCSS}
\eeqn
with $\ell$ the AdS radius, $G$ Newton's constant, dimensionless chiral functions $\bar L(x^+)$ and $\partial_+ \bar{P}(x^+)$ (which is also periodic) and $\Delta$ a constant. The conformal boundary is located at $\rho \to \infty$ and $x^\pm={ t\over \ell} \pm \phi$ with $\phi \sim \phi + 2\pi$.

BTZ black holes with mass $M$ and angular momentum $J$ are included in this family of metrics for vanishing $\partial_{+} \bar{P}$ and $\bar{L}(x^+)=\bar{\Delta}$, with $\ell M = \Delta + \bar{\Delta}$ and $J = \Delta - \bar{\Delta}$. Global AdS$_3$ is recovered as usual for $M=-1/8G$ and $J=0$.

Infinitesimal transformations leaving \eqref{SolCSS} invariant are given by asymptotic Killing vectors (AKVs)
\begin{equation}
\xi=
\epsilon \partial_+ + \left( \sigma +\frac{\ell^2}{2\rho^2}\partial^2_+\epsilon\right) \partial_-  -\frac{\rho}{2}\partial_+\epsilon \partial_\rho    +  \mathcal{O}(\ell^4/\rho^4).
\label{AKV}
\end{equation} 
and depend on two chiral arbitrary functions $\epsilon(x^+)$ and $\sigma(x^+)$. Expanding in Fourier modes,
one finds the conserved charges 
\beqn
{\mathcal{L}_m} := Q_{\epsilon = e^{i m x^+}} &=& \frac{1}{2 \pi} \int_{0}^{2 \pi} \D \phi \ e^{i m x^+}\left(\bar{L}-\Delta\left(\partial_+ \bar{P}\right)^{2}\right), \nonumber\\ 
{\mathcal{P}_m} := Q_{\sigma = e^{i m x^+}} &=& \frac{1}{2 \pi} \int_{0}^{2 \pi} \D \phi \ e^{i m x^+} \left( \Delta + 2 \Delta \partial_{+} \bar{P}\right),
\label{2.9}
\eeqn 
satisfying a Virasoro-Ka\v{c}-Moody algebra:
\beqn
i\left\{{\mathcal{L}}_{m}, {\mathcal{L}}_{n}\right\} &=& (m-n) {\mathcal{L}}_{m+n}+\frac{c_{R}}{12} m^{3} \delta_{m,-n}, \nonumber\\
i\left\{\mathcal{L}_{m}, \mathcal{P}_{n}\right\} &=& -n \mathcal{P}_{m+n},\nonumber\\
i\left\{{\mathcal{P}}_{m}, {\mathcal{P}}_{n}\right\} &=&\frac{k_{K M}}{2} m \delta_{m,-n}.
\label{CSSalgebra}
\eeqn
with
\begin{equation}
c_{R}=\frac{3 \ell}{2 G},  \qquad k_{K M}=-4 \Delta = -4 {\mathcal{P}}_{0}.
\label{2.12}
\end{equation}
Notice the unusual fact that the level is charge-dependent. A WCFT with this symmetry algebra is said to be in {\itshape quadratic ensemble}, and can be brought to canonical form either using state-dependent AKVs or a non-local redefinition of the charges \cite{Detournay:2012pc, Apolo:2018eky, Apolo:2020qjm}. In that case, the level is a negative constant. While this indicates non-unitarity of the theory, it appears to be a feature of holographic WCFTs and does not prevent, for instance, to apply techniques of the modular bootstrap to constrain the spectrum of the theory \cite{Apolo:2018eky, Chaturvedi:2018uov}.

 \section{CSS in TMG}

Topologically massive gravity \cite{Deser:1981wh} is described by the following three-dimensional action
\begin{equation}
I_{T M G}=\frac{1}{16 \pi G}\int_{M} \D^{3} x \sqrt{-g}(R-2 \Lambda)+\frac{1}{\mu} I_{C S}.
\label{Itmg}
\end{equation} 
The gravitational Chern-Simons term $ I_{C S}$ is given by
\begin{equation}
I_{C S}=\frac{1}{32 \pi G} \int_{M} \D^{3} x \sqrt{-g} \epsilon^{\lambda \mu \nu} \Gamma_{\lambda \sigma}^{\alpha}\left(\partial_{\mu} \Gamma_{\alpha \nu}^{\sigma}+\frac{2}{3} \Gamma_{\mu \tau}^{\sigma} \Gamma_{\nu \alpha}^{\tau}\right),
\label{3.2}
\end{equation} 
where $\Lambda = \pm 1/ \ell^2$ is the cosmological constant and $\mu$ the Chern-Simons coupling. Here we consider $\Lambda = - 1/ \ell^2$. Notice that our procedure for computing charges does not depend on the boundary terms needed to held a well-defined variational principle \cite{Barnich:2001jy,Barnich:2007bf,Compere:2018aar}.

Introducing the Einstein tensor
\begin{equation}
\mathcal{G}_{\mu \nu} \equiv R_{\mu \nu}-\frac{1}{2} g_{\mu \nu} R+\Lambda g_{\mu \nu}
\label{3.4}
\end{equation} 
and the Hodge-dualized Cotton tensor 
\begin{equation}
C_{\mu \nu} \equiv \epsilon_{\mu}^{\; \; \alpha \beta} \nabla_{\alpha}\left(R_{\beta \nu}-\frac{1}{4} g_{\beta \nu} R\right),
\label{3.5}
\end{equation}
the equations of motion are
\begin{equation}
\mathcal{G}_{\mu \nu}+\frac{1}{\mu} C_{\mu \nu}=0.
\label{3.3}
\end{equation} 
 Any solution of Einstein gravity with negative cosmological constant is automatically a solution of TMG. In particular, the metric \eqref{SolCSS} is a solution of \eqref{3.3}. 

\subsection{Charge algebra}

The conserved charges in TMG are modified with respect to their expressions in Einstein gravity and will be denoted with a tilde. The infinitesimal charge difference between two metrics $\bar g$ and $g = \bar g + \delta g$ associated to an AKV $\xi$ is given by
\begin{equation}
 \delta \tilde Q_{\xi}[g ; \bar{g}]=\int_{C} \sqrt{-g} \tilde k_{\xi}^{\mu \nu}[\delta g ; g] \epsilon_{\mu \nu \alpha} \D x^{\alpha},
\label{chargeTMG}
\end{equation} 
where the expression of the 1-form $\tilde k_{\xi}^{\mu \nu}$ can be found in \cite{Compere:2008cv} (see also eq. (142) of \cite{Detournay:2012pc}) and $C$ is a fixed-time contour at the AdS boundary. The Virasoro-Ka\v{c}-Moody charges then become
\beqn
\tilde{\mathcal{L}}_m &=& \frac{1}{2 \pi  \mu  \ell }\int_0^{2\pi } \D \phi  \ e^{imx^+} \left[(\mu  \ell +1) \bar{L}-   (\mu  \ell -1) \Delta(\partial_+\bar{P})^2\right],\nonumber \label{TMGG1}\\
\tilde{\mathcal{P}}_m &=& \frac{1}{{2 \pi  \mu  \ell }}\int_0^{2\pi } \D \phi  \ e^{imx^+} (\mu \ell - 1)(2\Delta \partial_+\bar{P} + \Delta)\label{TMGG2}
\eeqn
satisfying the same algebra as \eqref{CSSalgebra} with modified central extensions given by
 \beqn
   \tilde{c}_R = \Big(1+{1 \over \mu \ell}\Big) c_R \quad, \quad \tilde{k}_{KM} =  \Big(1-{1 \over \mu \ell}\Big) k_{KM}.
   \label{central}
\eeqn
We will hereafter focus on the signs of $\tilde{c}_R$ and $\tilde{k}_{KM}$ pertaining to holographic WCFTs, hence we will consider $\mu \ell \leq -1$ or $\mu \ell \geq 1$.

\subsection{Special points}

We see from \eqref{TMGG1} and \eqref{central} that at the point $\mu \ell  = 1$, the asymptotic symmetry group reduces to a chiral Virasoro algebra, because the Ka\v{c}-Moody charges vanish (i.e. become trivial).

To study the point $\mu \ell= -1$, we perform a redefinition of the arbitrary function $\sigma$
\beq
\sigma(x^+) \rightarrow \sigma(x^+)+ \sigma(x^+) \partial_+ \bar{P}(x^+).\label{red}
\eeq
Under this shift the charges become
\beqn
\tilde{\mathcal{L}}_m &=& \frac{1}{2 \pi} \Big(1+\frac{1}{  \mu  \ell}\Big) \int_0^{2\pi} \D\phi   e^{imx^+}   \bar{L},\nonumber\label{TMGG1fr}\\
\tilde{\mathcal{P}}_m &=& \frac{1}{2 \pi} \Big(1-\frac{1}{  \mu  \ell}\Big)  \int_0^{2\pi} \D\phi  e^{imx^+}\left(2 \Delta  \partial_+ \bar{P}+ \Delta \right)
\eeqn
and the mixed commutator $i\left\{\mathcal{L}_{m}, \mathcal{P}_{n}\right\}$ in \eqref{CSSalgebra} vanishes. The action of our redefinition has been to disentangle the Virasoro sector of the algebra from the Ka\v{c}-Moody one, for any value of $\mu$. Now, at $\mu \ell=-1$ the Virasoro generators \eqref{TMGG1fr} vanish identically as does the central charge, and the total algebra is simply given by a pure $\hat u(1)$ Ka\v{c}-Moody:
\beq
i\{\tilde{\mathcal{P}}_m, \tilde{\mathcal{P}}_n\} =  k_{K M}m \delta_{m,-n}.\label{u1TMG}
\eeq
This shows that there exists a particular value of the Chern-Simons coupling ($\mu \ell=-1$) where the total asymptotic algebra of $3$-dimensional TMG with negative cosmological constant is a $\hat u(1)$ Ka\v{c}-Moody current algebra.

\section{Black Holes}

The BTZ metric in ADM form is
\beqn
\D s^{2} = -N^{2} \D t^{2}+\frac{\D r^{2}}{N^{2}}+r^{2}\left(N^{\phi}\D t+\D \phi\right)^{2},
\label{BTZ}
\eeqn 
with
\beq
N^{2} =-8 G M+\frac{r^{2}}{\ell^{2}}+\frac{16 G^{2} J^{2}}{r^{2}} = \frac{\left(r^{2}-r_{+}^{2}\right)\left(r^{2}-r_{-}^{2}\right)}{r^{2} \ell^{2}}
\eeq
and
\beq
N^{\phi} = -\frac{4 G J}{r^{2}}.
\eeq
The black hole horizons are located at
\beqn
r_{\pm} = \sqrt{2 G \ell(\ell M+J)} \pm \sqrt{2 G \ell(\ell M-J)}. \label{BTZhorizon}
\eeqn

The black hole mass $\tilde{M} = Q_{\partial_t}$ and angular momentum $\tilde{J} = Q_{-\partial_\phi}$ in TMG depart from their values $M$ and $J$ in pure gravity and are given by
\beqn
\ell \tilde{M} &=& \ell M-\frac{J}{\mu \ell},\label{MassTMG}\\ 
\tilde{J} &=& J- \frac{M}{\mu} \label{AngularTMG}.
\eeqn 
Absence of naked singularities imposes $\ell M\geq |J|$. Therefore, in order to have positive energies $\ell \tilde M>0$, we need to consider $|\mu| \geq {1\over \ell}$, which is the condition we had already obtained under \eqref{central}.
At the special points $\mu \ell  = \pm 1$ we have the extremality conditions
\beqn
\ell \tilde{M} = \mp \tilde{J}.
\eeqn

The BTZ entropy in TMG has been computed in \cite{Kraus:2005zm, Solodukhin:2005ah, Tachikawa:2006sz, Bouchareb:2007yx} resulting in
\beqn
\tilde{S} =\frac{\pi r_{+}}{2 G}- \frac{\pi r_{-}}{2 G \mu \ell}. \label{EntropyTMG}
\eeqn 
We expect this to be reproduced by counting the degeneracy of states in the dual WCFT. In the quadratic ensemble, the warped Cardy formula takes the form \cite{Detournay:2012pc}
\beqn
S_{\text{WCFT}} = 4 \pi \sqrt{-\tilde{\mathcal{P}}_{0}^{v a c} \tilde{\mathcal{P}}_{0}}+4 \pi \sqrt{-\tilde{\mathcal{L}}_{0}^{v a c} \tilde{\mathcal{L}}_{0}}. \label{WCFTEntr}
\eeqn 
In this expression the subscript $vac$ refers to the charges of the vacuum. Here, the vacuum is global AdS$_3$, whose charges are $M = -1/8G$ and $J=0$.  
For the BTZ black hole, the zero modes in \eqref{TMGG1fr}
are given by
\beqn
\tilde{\mathcal{L}}_0 = \Big(1+ \frac{1}{\mu \ell}\Big)  \left(\frac{\ell M-J}{2}\right), \label{ZeroL} 
\eeqn
and
\beqn
\tilde{\mathcal{P}}_0=  \Big(1-\frac{1}{\mu \ell}\Big) \left(\frac{\ell M+J}{2}\right) \label{ZeroP}.
\eeqn 
Plugging this in \eqref{WCFTEntr} using \eqref{BTZhorizon}, one finds $\tilde{S} = S_{\text{WCFT}}$\footnote{This had been shown for any diffeomorphism-invariant higher curvature theory in \cite{Zwikel:2016smm}}. At the special points $\mu \ell  = \pm 1$, one observes that the BTZ black hole entropy is reproduced from the contributions of a chiral Virasoro or $\hat u(1)$ Ka\v{c}-Moody current algebra only.

\section{Gravitons}

In this section we solve the spin-$2$ linearized perturbation around the AdS$_3$ background in TMG.  While we begin following closely \cite{Li:2008dq}, we then require different conditions on the perturbations -- due to the different asymptotic symmetries under consideration. We end up with numerous perturbations solving the linearized TMG equations of motion of which we keep the ones with finite energy and regular at the origin. 

We consider AdS$_3$ in global coordinates\footnote{Abusing notation, we call again the holographic coordinate $\rho$, although it is the logarithm of the Fefferman-Graham coordinate (also spelled $\rho$). The latter has dimension length while the former is dimensionless.}
\begin{eqnarray}
\D s^2  &=&-\frac{1}{4} \ell^2 \left(-4 \D\rho ^2+\D x^{+2}+2 \D x^+ \D x^- \cosh (2 \rho )+ \D x^{-2}\right)\nonumber \\ &:=& \bar g_{\mu\nu}\D x^\mu\D x^\nu .
\label{eq:1}
\end{eqnarray} 

The isometry group of the metric \eqref{eq:1} is $S L(2, \mathbb{R})_{L} \times S L(2, \mathbb{R})_{R}$ with generators $\bar{L}_{0,\pm 1}$ and $L_{0,\pm 1}$ respectively. We will single out a $U(1)\times S L(2, \mathbb{R})_{R} $ subalgebra compatible with the CSS boundary conditions to classify the perturbations, the $U(1)$ factor being generated by $P_0=i \pa_- = \bar{L}_0$ and the relevant $S L(2, \mathbb{R})_{R}$ generators given by
\beqn
L_{0}  &=& i \partial_{+}, \\
 L_{1} &=& i e^{i x^+}\left[\frac{\cosh 2 \rho}{\sinh 2 \rho} \partial_{+}-\frac{1}{\sinh 2 \rho} \partial_{-}-\frac{i}{2} \partial_{\rho}\right].
\label{eq:3}
\eeqn 

We write linearized excitations around the AdS$_3$ background metric $\bar{g}$ as
\begin{equation}
g_{\mu \nu}=\bar{g}_{\mu \nu}+h_{\mu \nu},
\label{eq:2.1}
\end{equation} 
with $h_{\mu \nu}$ a small perturbation. The linearized equations of motion in TMG are 
\begin{equation}
\mathcal{G}_{\mu \nu}^{(1)}+\frac{1}{\mu} C_{\mu \nu}^{(1)}=0,
\label{eq:2.2}
\end{equation} 
where explicit expressions for the linearized Einstein and Cotton tensors can be found in \cite{Li:2008dq, Maloney:2009ck}.
In transverse and traceless gauge
\begin{equation}
\bar{\nabla}_\mu h^\mu_{\; \; \nu}=h=0,
\label{eq:2.9}
\end{equation}
the equations of motion can be recast as \cite{Li:2008dq, Maloney:2009ck}
\begin{equation}
\left(\bar{\nabla}^{2}+\frac{2}{\ell^{2}}\right)\left(h_{\mu \nu}+\frac{1}{\mu} \ve_{\mu}{}^{\alpha \beta} \bar{\nabla}_{\alpha} h_{\beta \nu}\right)=0.
\label{eq:2.13}
\end{equation}

We want to use the $U(1) \times S L(2, \mathbb{R})_{R} $ algebra to classify linear perturbations. 
Consider thus primary states with weight $(h, p)$:
\begin{equation}
L_{0}\left|h_{\mu \nu}\right\rangle=h\left|h_{\mu \nu}\right\rangle, \quad P_{0}\left|h_{\mu \nu}\right\rangle=p \left|h_{\mu \nu}\right\rangle,
\label{eq:2.22}
\end{equation}
which implies
\begin{equation}
h_{\mu \nu}=e^{-i (h x^++p x^-)} F_{\mu \nu}(\rho).
\label{eq:2.23}
\end{equation}

The transverse, traceless, and highest-weight condition $L_1 \left|h_{\mu \nu}\right\rangle = \text{Lie}_{L_1}\left|h_{\mu \nu}\right\rangle = 0$ will strongly constrain the form of $F_{\mu \nu}(\rho)$, whose components will depend on $p$, $h$, and a set of integration constants. Inserting the result in the equations of motion, and requiring finiteness of the linearized energy of the real part of the perturbation (eq. (69) of \cite{Li:2008dq}), regularity at the origin and to satisfy the CSS boundary conditions singles out 3 solutions:

\begin{itemize}
\item Massive mode : $h=\frac{1}{2}(3-\mu \ell)$, $p=-\frac{1}{2}(1+\mu \ell) $ and $\beta = 0$ hereunder, with energy 
\beq
E =-\frac{ \alpha^2 (\mu  \ell-1)^2 (\mu  \ell+1)}{256  G \mu  \ell^6 (4-2 \mu  \ell)}
\eeq for $\mu \ell < 2$ and diverge for $\mu \ell \geq 2$. 
\item Right graviton mode : $h=2 $, $p=0$ and $\beta = 0$ hereunder, with energy 
\beq
E=  \frac{\alpha^2 (\mu  \ell +1)}{384 G \mu  \ell^6}.
\eeq 
\item Right photon mode :  $h=1 $, $p=0$ and $\alpha = 0$ hereunder, with energy \footnote{Using the same terminology as \cite{Compere:2008cv}, we call the photon the mode with $h=1$.}
\beq
E= \frac{ \beta^2 (\mu  \ell-1)}{32  G \mu  \ell^6},
\eeq
\end{itemize}

with explicit wavefunctions

\beqn
F_{++} &=& \frac{1}{4} \cosh ^{4-2h}\rho  \tanh ^{p-h}\rho \left(4 \beta \tanh ^2\rho +\alpha \tanh ^4\rho\right), \label{F++} \\
 F_{+-}&=& \frac{1}{2} \cosh^{2(1-h)}\rho\tanh^{p-h}\rho  \left(\beta \tanh ^2\rho \right),  \\
 F_{+ \rho}&=&  \frac{i}{32}  \sinh^{-1}\rho \cosh^{-(1+2h)}\rho  \tanh ^{p-h}\rho [4 \cosh 2 \rho (2 \beta-\alpha)\nonumber\\&&-8 \beta + 3 \alpha+\alpha(\cosh 4 \rho )] ,\\
  F_{--}&=&0, \\
   F_{-\rho}&=& -\frac{i}{4}  \cosh^{-1}\rho \sinh ^{-1}\rho \sinh^{-h}2\rho  \tanh ^{p-h}\rho  
   \nonumber\\
   &&[\sinh ^h2 \rho  \cosh^{-2h}\rho   ((-\beta) \cosh 2 \rho +\beta)],\\
   F_{\rho \rho}&=& \sinh ^{-h-2}2 \rho  \tanh ^{p-h}\rho[\cosh ^{4-2h}\rho \sinh ^h2 \rho \nonumber\\ &&\left((4 \beta-\alpha) \tanh ^4\rho \right)]
    , \label{Frr}
\eeqn

The first two modes were present in \cite{Li:2008dq}, and their energies coincide with (70)-(71) of that reference\footnote{Remember that left and right are flipped and that $\mu_{\text{here}} = - \mu_{\text{there}}$ because of a different convention in the $\ve$ symbol.  Our convention is $\varepsilon_{+-\rho}=-1$.}. There is no Left graviton mode with $h=0$ and $p=2$, as it is excluded by the CSS boundary conditions. Instead, there is a new solution, the Right photon mode which satisfies CSS but not Brown-Henneaux.  

For $\mu \ell < -1$, the massive mode has a negative energy while the graviton and the photon mode have a positive energy. At the special point $\mu \ell = -1$  the massive mode and the graviton have a zero energy while the photon has a positive energy. At the special point $\mu \ell =1$ the massive mode and the photon have zero energy while the graviton has a positive energy. For $\mu \ell  > 1$  the massive mode has a negative energy while the graviton and the photon mode both have a positive energy.  We thus see that the two special points $\mu \ell  = \pm 1$ allow to avoid negative energy, while the right moving graviton or photon carry no energy.

\section{Summary and further developments}

In this work we investigated CSS boundary conditions in TMG for $|\mu | \ell \geq 1$ (for which BTZ black holes have a positive energy)
and noticed two special points in the space of couplings.

At $\mu \ell = 1$:
\begin{itemize}
\item 
The $\hat u(1)$ Ka\v{c}-Moody charges and level vanish;

\item
BTZ black holes have positive energy, an entropy reproduced by a chiral half of the Cardy formula and $\ell \tilde{M} = - \tilde{J}$;

\item
The massive graviton and boundary photon acquire vanishing energy, while the boundary graviton has positive energy.
\end{itemize}

At $\mu \ell = -1$:
\begin{itemize}
\item 
The Virasoro generators and central charge vanish;

\item
BTZ black holes have positive energy, an entropy reproduced by the $\hat u(1)$ contribution of the Warped Cardy formula and $\ell \tilde{M} = \tilde{J}$;

\item
The massive and boundary gravitons acquire vanishing energy, while the boundary photon has positive energy.
\end{itemize}

This suggests that TMG with CSS boundary conditions at these two points might be a stable and consistent gravity theory dual either to a holomorphic CFT, or a theory with 
$\hat u(1)$ affine symmetry.

In this work we have considered a phase space consisting of Einstein solutions. It is known however that there exist a large variety of solutions to TMG which are not Einstein \cite{Chow:2009km}. CSS should therefore be generalized to TMG, in the spirit of \cite{Henneaux:2010fy}. In particular, logarithmic solutions are likely to arise at the special points. This might in turn lead to the definition of ``logarithmic WCFT" that could be dual to a relaxation of CSS in TMG at the special points. A positive energy theorem should also be proven. A preliminary step would be to show that the only stationary, axially symmetric solutions of the theory are BTZ black holes. This is true for chiral gravity \cite{Maloney:2009ck}, even though non-Einstein time-dependent solutions do exist \cite{Compere:2010xu}.
It would also be interesting to evaluate the Euclidean sum over geometries and determine what type of partition it can be identified with.

\acknowledgments

We thank Alessio Caddeo, Marc Henneaux, Wout Merbis, Blagoje Oblak, Antonin Rovai, Wei Song and Bayram Tekin for valuable discussions.
The work of LC is supported by the ERC Advanced Grant ``High-Spin-Grav". SD is a Research Associate of the Fonds de la Recherche Scientifique F.R.S.-FNRS (Belgium). SD was supported in part by IISN -- Belgium (convention 4.4503.15) and benefited from the support of the Solvay Family. SD acknowledges support of the Fonds de la Recherche Scientifique F.R.S.-FNRS (Belgium) through the CDR project C 60/5 - CDR/OL ``Horizon holography : black holes and field theories" (2020-2022).  SD thanks the KITP for its hospitality during the completion of this work. This research was supported in part by the National Science Foundation under Grant No. NSF PHY-1748958.  The work of AS is supported by  the ``Fonds pour la Formation et la Recherche dans l'Industrie et dans l'Agriculture", FRIA (Belgium).

\providecommand{\href}[2]{#2}\begingroup\raggedright\endgroup

 
\end{document}